# Standing wave approach in the theory of x-ray magnetic reflectivity

M. A. Andreeva[1], P. A. Baulin[1], Yu. L. Repchenko[2]

[1] *Faculty of Physics, M.V. Lomonosov Moscow State University, 119991, Moscow, Russia*

[2] *National Research Centre "Kurchatov Institute", Pl. Kurchatova 1, 123182, Moscow, Russia*

We have developed the extension of the exact x-ray resonant magnetic reflectivity theory taking into account the small value of the magnetic terms in the x-ray susceptibility tensor. We have shown that squared standing waves (forth powder of the total electric field module) determine the output of the magnetic addition to the total reflectivity from a magnetic multilayer. The approach gives the easily understandable results especially for the reflectivity part with the rotated polarization and the obtained relationship can essentially speed up the calculation of the asymmetry ratio of the magnetic reflectivity relative the sign of the circular polarization in the incident radiation. The connection of the magnetic contribution with the standing wave structure in the magnetic multilayer means that the magnetic reflectivity with rotated polarization possesses higher sensitivity to the depth profiles of magnetization and magnetic elements depth distribution than the secondary radiation registration at the specular reflection condition.

## I. INTRODUCTION

Polarization properties of radiation absorbed or scattered by magnetized samples plays more and more important role in the magnetic property investigations with synchrotron radiation. The modern synchrotron lines produce or create x-rays of any desired polarization state and the polarization dependent absorption or scattering near the x-ray absorption edges (XMCD, XMLD, XMND, M$\chi$D, DAFS, XRMR) has become the basis of the extremely effective methods for magnetic or structure investigations [1-7]. Polarization analysis in the nonresonant magnetic x-ray scattering has been proved to be the very effective tool for separation of the charge and magnetic scatterings (revealing e.g. the difference in the charge and magnetic periodicity [8]), and supplies as well the spin and orbital magnetic moment determination [9-17]. Magneto-optical measurements

(Faraday and Voigt Effects) also utilized the polarization analysis of the transmitted radiation [18-22]. In the resonant x-ray diffraction the selection of the $\sigma \to \pi'$ channel supplies the possibility to enhance the fine effects like quadrupole transition contributions, orbital ordering, Dzyaloshinskii-Moriya interaction and interference of the quadrupole resonant and nonresonant scattering amplitudes in the structurally forbidden reflections [23-25].

X-ray resonant magnetic reflectivity (XRMR) gives unique elemental and spatially selective information about magnetic ordering in multilayer films (see e.g. [26-40]). In these experiments two circular or two linear polarizations of synchrotron radiation are used and the asymmetry ratio relative the polarization state of the incident radiation is measured. The change of polarization of the reflected radiation (x-ray Kerr rotation) has been directly measured with soft x-rays only in a few works [41-43]. Ellipsometry measurements have mainly been conducted in the visible range [44] but the usefulness of the ellipsometric investigations in the x-ray region have not been focused on until now. From the general point of view the polarization analysis of reflected radiation should be the source of the valuable information supplemented to the asymmetry ratio data. Moreover, as it pointed out in the papers [45-46], following the development of femtosecond light sources such as synchrotron radiation (SR) sources using a laser slicing technique and the free electron lasers, the x-ray Kerr rotation is becoming increasingly important, particularly for temporal domain measurements in the subpicosecond timescale.

Polarization dependence of the reflectivity is obtained by exact calculations of the reflectivity amplitude. Magnetic scattering, being significant near the absorption edges of magnetic atoms, radically complicates the theory of reflectivity, because the x-ray susceptibility of a medium becomes a tensor in the presence of the magnetic scattering. The reflectivity theory from anisotropic (magnetic) multilayers was developed in [31, 47-50] based on the eigen-wave formalism or by using the method of the 4x4 propagation matrices in [44, 51-55]. The application of both algorithms for interpreting real experimental data is rather time consuming; therefore, simplifying the calculations is an urgent problem. In particular, the analytical expressions were obtained for the integral propagation 4x4-matrix [53, 56-58] using some simplifications which are not always valid [59]. An interesting approach, namely a combination of dynamical (for the isotropic scattering part) and kinematical (for the magnetic scattering part) approximations, was used in [38] to interpret the reflection curves near the Dy $M_5$ absorption edge from a dysprosium film with helicoid magnetic ordering. A kinematic scalar approximation was used in [32, 33] when interpreting the asymmetry ratio of the XRMR spectra to obtain the depth-profiles of the spin polarization in 5*d* electron shells of cerium and lanthanum in [Ce/Fe]$_n$ and [La/Fe]$_n$ multilayers. The polarization asymmetry ratio is generally very small and extremely sensitive to calculation errors. Correct interpretation of the experimental data is very important in such studies. It has been shown in [60] that the kinematic theory of reflectivity is applicable at the angles far enough from the total external reflection region but in some cases the complex polarization dependence of the propagating radiation leads to the wrong results even at rather large angles of incidence.

In this paper we show that the angular (and energy) dependence of the dichroic part of the magnetic contribution to the x-ray reflectivity is connected with the squared standing waves of the radiation inside the reflecting sample. It is mostly pronounced in the region of the total external reflection and at other angles where the reflectivity is high enough. This founding explains the specific features of the angular dependence of the reflectivity with the rotated polarization and reveals the enhanced depth selectivity of this part of reflectivity. In addition the presented approach allows to simplify the calculation of the reflectivity from magnetic multilayers because it implies the substitution of the complicated calculations with 4x4 propagation matrices (or eigen waves) by the simple Parratt algorithm for the isotropic part of multilayer and the integration of the magnetic part with "weight" of squared standing waves for the rotated part of the reflectivity. We present some test calculations to verify this approximation under certain conditions and show the first experimental results demonstrating the validity of the predicted angular dependence for the rotated part of reflectivity.

## II. X-RAY REFLECTIVITY FROM ULTRATHIN LAYER

Reflectivity from ultrathin layer in the case of scalar susceptibility $\chi(z)$ can be easily obtained using the propagation matrix method [44, 51]. In the case of planar structures we use the tangential components of the electric $E_t = |\mathbf{q} \times \mathbf{E}|$ and magnetic $H_t = -|\mathbf{q} \times \mathbf{q} \times \mathbf{H}|$ field of the plane electromagnetic wave $\sim \exp(ikr - i\omega t)$ (**q** is the unit vector normal to the surface) to describe the variations of the radiation field amplitudes along depth z, so the Maxwell equations come to the differential equation [44, 61]:

$$\frac{d}{dz}\begin{pmatrix} E_t(z) \\ H_t(z) \end{pmatrix} = ik\hat{M}(z)\begin{pmatrix} E_t(z) \\ H_t(z) \end{pmatrix}, \qquad (1)$$

where $k = \dfrac{\omega}{c} = \dfrac{2\pi}{\lambda}$ is the wave vector of the incident electromagnetic wave and $\hat{M}$ is the differential propagation matrix which for e.g. $\sigma$-polarization of radiation takes a form:

$$\hat{M}^\sigma(z) = \begin{pmatrix} 0 & 1 \\ \sin^2\theta + \chi(z) & 0 \end{pmatrix}, \qquad (2)$$

where $\theta$ is the grazing angle of incidence of the plane wave and $\chi(z)$ is the susceptibility of the layered sample. The integral propagation matrix in a layer of thickness $d$ and scalar susceptibility $\chi(z) = const$ can be easily calculated

$$\hat{L}(d) = \begin{pmatrix} l_{11} & l_{12} \\ l_{21} & l_{22} \end{pmatrix} = \exp(ikd\,\hat{M}). \qquad (3)$$

It connects the radiation field amplitudes on the top and bottom boundaries of the layer

$$\begin{pmatrix} E_T \\ \eta_T E_T \end{pmatrix} = \hat{L}(d) \begin{pmatrix} E_0 + E_R \\ \eta_0(E_0 - E_R) \end{pmatrix}, \qquad (4)$$

where the subscripts $0, R, T$ designate the incident, reflected and transmitted waves, $\eta_0 = \sqrt{\sin^2\theta + \chi_0}$ and $\eta_T = \sqrt{\sin^2\theta + \chi_T}$ are the normal components of the wave vectors in units of $\dfrac{\omega}{c}$ in the outer medium and in the substrate. If on the outside of the layer $\chi_0 = 0$ and $\chi_T = 0$, then $\eta_0 = \eta_T = \sin\theta$. In (4) we have taken into account that in the case of the $\sigma$-polarization $E_t = |\mathbf{E}| = E$ and the magnetic field of radiation $H_{tj} = \pm\eta_j E_j$ (the $\pm$ sign refers to the waves in the direct and opposite directions, $j = 0, R, T$). The system of two equations (4) gives the reflection amplitude

$$r = E_R / E_0 = \dfrac{(\eta_0 l_{22} + l_{21}) - \eta_T(\eta_0 l_{12} + l_{11})}{(\eta_0 l_{22} - l_{21}) - \eta_T(\eta_0 l_{12} - l_{11})}. \qquad (5)$$

Supposing the thickness of the layer very small, the matrix exponential in (3) can be approximated by the expression ($\eta = \sqrt{\sin^2\theta + \chi}$):

$$\exp(ikd\hat{M}) \cong 1 + ikd\hat{M} = \begin{pmatrix} 1 & ikd \\ ikd\,\eta^2 & 1 \end{pmatrix}. \qquad (6)$$

If the layer is placed in vacuum ($\chi_0 = \chi_T = 0$), substituting the elements of the integral propagation matrix from (6) into (5), the reflection amplitude from ultrathin layer takes a well known form:

$$r \cong \frac{\sin\theta + ikd\,\eta^2 - \sin^2\theta\,ikd - \sin\theta}{\sin\theta - ikd\,\eta^2 - \sin^2\theta\,ikd + \sin\theta} \cong \frac{ikd\,\chi}{2\sin\theta} \qquad (7)$$

The same expression can be obtained from the Parratt recurrent equations. The Parratt method allows to take into account the modification of the expression (7) in the case when the layer is placed under some reflecting substrate. In this case the Parratt algorithm gives for the reflectivity from the whole system the following expression:

$$R^{tot} = \frac{r_{01} + \tilde{R}e^{2i\varphi}}{1 + r_{01}\tilde{R}e^{2i\varphi}}, \qquad (8)$$

where $r_{01}$ is the Fresnel amplitude of the single reflection from the film surface

$$r_{01} = \frac{\sin\theta - \eta}{\sin\theta + \eta}, \qquad (9)$$

$$2i\varphi = ik\eta d \qquad (10)$$

is the phase incursion in the film and

$$\tilde{R} = \frac{r_{10} + R}{1 + r_{10}R}, \qquad (11)$$

is the reflectivity from the bottom boundary of the film, $R = R^{substr}e^{iQH}$, $Q = \frac{4\pi}{\lambda}\sin\theta$, $H$ is the distance from substrate to the film, $R^{substr}$ is the reflectivity from the substrate (see Fig. 1).

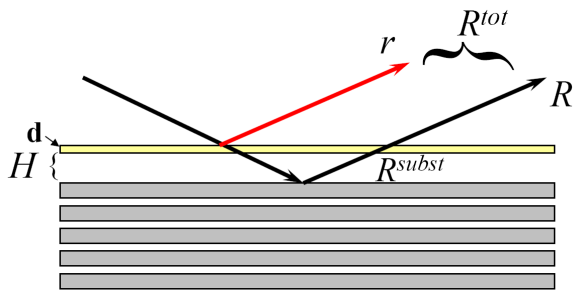

*Fig. 1. The illustration to the modification of the reflectivity from ultrathin layer in the presence of the reflecting substrate.*

Using for an ultrathin film the expansion $e^{2i\varphi} \approx 1 + 2i\varphi$ and keeping only the first degree of $2i\varphi$, we transforms (8) to:

$$R^{tot} \approx \frac{r_{01}+\tilde{R}}{1+r_{01}\tilde{R}} + \frac{\tilde{R}(1-r_{01}^2)}{(1+r_{01}\tilde{R})^2} 2i\phi. \tag{12}$$

If $\tilde{R} = r_{10} = -r_{01}$ (film in vacuum), we easily get (7):

$$R^{tot} \approx -\frac{r_{01}}{(1-r_{01}^2)} 2i\phi = \frac{\eta^2 - \sin^2\theta}{4\sin\theta\,\eta} i2k\eta d = \frac{ikd\,\chi}{2\sin\theta} \cong r. \tag{13}$$

Substituting (11) into (12) we get for the first term

$$\frac{r_{01}+\dfrac{r_{10}+R}{1+r_{10}R}}{1+r_{01}\dfrac{r_{10}+R}{1+r_{10}R}} = \frac{r_{01}(1+r_{10}R)+(r_{10}+R)}{(1+r_{10}R)+r_{01}(r_{10}+R)} = R \tag{14}$$

and for the second term

$$\frac{\dfrac{r_{10}+R}{1+r_{10}R}(1-r_{01}^2)}{(1+r_{01}\dfrac{r_{10}+R}{1+r_{10}R})^2} 2i\phi = \frac{(r_{10}+R)(1+r_{10}R)}{(1-r_{01}^2)} 2i\phi =$$

$$= \frac{(r_{10}+R+r_{10}r_{10}R+r_{10}R^2+2r_{10}R-2r_{10}R)}{(1-r_{01}^2)} 2i\phi = \tag{14}$$

$$= R\frac{(1+r_{01})}{(1-r_{01})} 2i\phi + \frac{r_{10}}{(1-r_{01}^2)}(1+R)^2 2i\phi$$

Taking into account (9), (10), (13) and the relation

$$\frac{(1+r_{01})}{(1-r_{01})} = \frac{(1-r_{10})}{(1+r_{10})} = \frac{\sin\theta}{\eta}, \tag{15}$$

finally we get:

$$R^{tot} \approx R + r(1+R)^2 + R\frac{\sin\theta}{\eta} ik\eta d = R^{substr}e^{iQH}(1+iQd) + r(1+R)^2 \approx$$
$$\approx R^{substr}e^{iQ(H+d)} + (1+R^{substr}e^{iQH})^2 r \tag{16}$$

The expression (16) means that the reflectivity amplitude from ultrathin layer $r$ placed above the reflecting substrate is modulated $r \to r'$

$$r' = (1+R^{substr}e^{iQH})^2 r = (E_{tot}(H))^2 r \tag{17}$$

by the squared total field amplitude $E_{tot}(H)$ (that is a standing wave amplitude) at the position $H$ of this thin layer. Two terms in (16) represent the reflectivity amplitude from the substrate (the first term) and from the ultrathin layer $r$, modulated by the squared standing wave, created by the interference of the incident and reflected from the substrate waves.

By some more lengthy calculations the expression for the modulated reflectivity from the ultrathin layer $r'$ in (16) can be generalized to the case when the ultrathin layer is placed inside the multilayer at some depth $z$:

$$r'(z) = T(z)T'(z)(1+R^{below}(z))^2 r = E_{tot}^{\;2}(z)\, r \,, \tag{18}$$

where $R^{below}(z)$ corresponds to the reflectivity amplitude from the part of the multilayer below the considered ultrathin layer, and the functions $T(z)T'(z)$ describe the transformations of the transmitted and outgoing waves during multiple reflections at all boundaries in the upper part of the multilayer:

$$T(z)T'(z) = e^{2i(\phi_1+\phi_2+\ldots+\phi_{j-1})} \times \frac{(1-r_1^2)(1-r_2^2)\ldots(1-r_{j-1}^2)}{(1+r_1 R_2 e^{2i\phi_1})^2 (1+r_2 R_3 e^{2i\phi_2})^2 \ldots (1+r_{j-1} R_j e^{2i\phi_{j-1}})^2}. \tag{19}$$

Taking into account the well-known Fresnel relations in each layer:

$$1-r_i^2 = (1-r_i)(1+r_i) = t_i t'_i, \quad t_i = \frac{2\eta_{i-1}}{\eta_{i-1}+\eta_i}, \quad t'_i = \frac{2\eta_i}{\eta_{i-1}+\eta_i}, \tag{20}$$

the two separate functions (each one for the waves in the forward and backward directions correspondingly) can be extracted from (19):

$$T(z) = e^{i(\phi_1+\phi_2+\ldots+\phi_{j-1})} \times \frac{(1+r_1)(1+r_2)\ldots(1+r_{j-1})}{(1+r_1 R_2 e^{2i\phi_1})(1+r_2 R_3 e^{2i\phi_2})\ldots(1+r_{j-1} R_j e^{2i\phi_{j-1}})} \tag{21}$$

and

$$T'(z) = T(z)\frac{\eta_j}{\eta_0}. \tag{22}$$

Omitting the multiple scatterings in the kinematical approximation, these functions are simplified to the well-known expression describing the phase incursion and as well the effects of the refraction and absorption for the transmitted and outgoing waves in the upper part of the multilayer:

$$T(z)T'(z) = e^{iQz + \frac{2\pi}{\lambda \sin \vartheta} \sum_{m=1}^{j-1} \chi_m d_m}. \tag{23}$$

The intensity of the total reflectivity includes the squared module of both terms and the interference term. The most interesting case happens when in the experiment it is possible to select only the second term $r'$. It takes place e.g. in the time domain nuclear resonant reflectivity when the resonance scattering gives the delayed signal and the resonant nuclei are placed only in thin film. In this case the time gating separates the response from the substrate and the film [63-66]. In this article we consider another way to separate the response from the substrate and the film, namely by the polarization analysis of the reflected radiation.

### III. ANYSOTROPIC ULTRATHIN LAYER

The reflectivity amplitude from an anisotropic layer can be considered similar to the isotropic case, namely by means of the propagation matrix. In the anisotropic case the propagation matrix is a 4x4 matrix and it describes the variation of the 2-dimential tangential vectors $\mathbf{H}_t = -\mathbf{q} \times \mathbf{q} \times \mathbf{H} = \begin{pmatrix} H_x \\ H_y \end{pmatrix}$ and $\mathbf{q} \times \mathbf{E} = \begin{pmatrix} -E_y \\ E_x \end{pmatrix}$ of the plane electromagnetic wave $\sim \exp(ikr - i\omega t)$ in a layered medium ($x$ axis is chosen perpendicular to the reflection plane, $y$ axis in the reflection plane – Fig. 2):

$$\frac{d}{dz}\begin{pmatrix} \mathbf{H}_t \\ \mathbf{q} \times \mathbf{E} \end{pmatrix} = ik\hat{M}\begin{pmatrix} \mathbf{H}_t \\ \mathbf{q} \times \mathbf{E} \end{pmatrix} \quad \text{or} \quad \frac{d}{dz}\begin{pmatrix} H_x(z) \\ H_y(z) \\ -E_y(z) \\ E_x(z) \end{pmatrix} = ik\,\hat{M}(z)\begin{pmatrix} H_x(z) \\ H_y(z) \\ -E_y(z) \\ E_x(z) \end{pmatrix}. \tag{24}$$

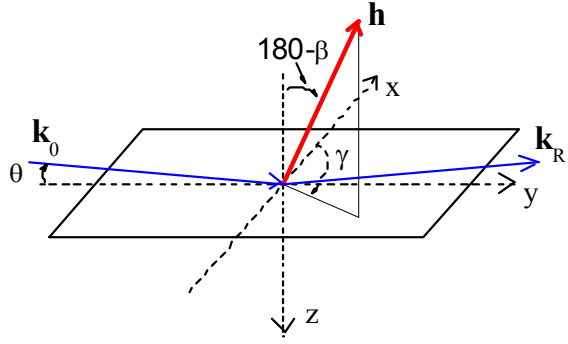

*Fig. 2. The used coordinate system.* $\mathbf{k}_0$ *and* $\mathbf{k}_R$ *are the wave vectors of the incident and reflected plane waves,* $\beta$ *and* $\gamma$ *are the polar and azimuth angles for the magnetization unit vector.*

In general case the matrix $\hat{M}(z)$ has been calculated in the textbook [44] by coordinate method and by covariant tensor method in [51-52]. If the material equations include just the tensor of the electric susceptibility $\hat{\chi}$ ($\mathbf{D} = \hat{\varepsilon}\mathbf{E} = (1+\hat{\chi})\mathbf{E}$, $\mathbf{B} = \mathbf{H}$), then $\hat{M}(z)$ has been presented in [51, 54] in the following form:

$$\hat{M} = \begin{pmatrix} \dfrac{1}{\varepsilon_q} \mathbf{q}^\times \hat{\varepsilon} \mathbf{q} \circ \mathbf{a} & \dfrac{1}{\varepsilon_q} \hat{\mathbf{I}} \hat{\tilde{\varepsilon}} \hat{\mathbf{I}} - \mathbf{b} \circ \mathbf{b} \\ \hat{\mathbf{I}} - \dfrac{1}{\varepsilon_q} \mathbf{a} \circ \mathbf{a} & -\dfrac{1}{\varepsilon_q} \mathbf{a} \circ \mathbf{q} \hat{\varepsilon} \mathbf{q}^\times \end{pmatrix} =$$

$$= \begin{pmatrix} -\dfrac{\chi_{yz}}{1+\chi_{zz}}\cos\theta & 0 & 1+\chi_{yy} - \dfrac{\chi_{zy}\chi_{yz}}{1+\chi_{zz}} & \dfrac{\chi_{yz}\chi_{zx}}{1+\chi_{zz}} - \chi_{yx} \\ \dfrac{\chi_{xz}}{1+\chi_{zz}}\cos\theta & 0 & \dfrac{\chi_{zy}\chi_{xz}}{1+\chi_{zz}} - \chi_{xy} & \chi_{xx} - \dfrac{\chi_{zx}\chi_{xz}}{1+\chi_{zz}} + \sin^2\theta \\ 1 - \dfrac{\cos^2\theta}{1+\chi_{zz}} & 0 & -\dfrac{\chi_{zy}}{1+\chi_{zz}}\cos\theta & \dfrac{\chi_{zx}}{1+\chi_{zz}}\cos\theta \\ 0 & 1 & 0 & 0 \end{pmatrix} \approx$$

$$\approx \begin{pmatrix} -\chi_{yz}\cos\theta & 0 & 1+\chi_{yy} & -\chi_{yx} \\ \chi_{xz}\cos\theta & 0 & -\chi_{xy} & \sin^2\theta + \chi_{xx} \\ \dfrac{\sin^2\theta + \chi_{zz}}{1+\chi_{zz}} & 0 & -\chi_{zy}\cos\theta & \chi_{zx}\cos\theta \\ 0 & 1 & 0 & 0 \end{pmatrix} \quad (25)$$

using the following notations: $\mathbf{b}$ is the unchanged tangential components of the all wave vectors in units of $\omega/c$, $\mathbf{a} = \mathbf{b}\times\mathbf{q}$, $|\mathbf{b}| = |\mathbf{a}| = \cos\theta$, $\hat{\mathbf{I}} = 1 - \mathbf{q}\circ\mathbf{q} = -(\mathbf{q}^\times)^2$ is the projective tensor, $\mathbf{q}^\times$ is a dual tensor performing the vector product, $\varepsilon_q = \mathbf{q}\hat{\varepsilon}\mathbf{q}$, $\tilde{\hat{\varepsilon}} = (\mathbf{q}^\times\hat{\varepsilon}\mathbf{q} \circ \mathbf{q}\hat{\varepsilon}\mathbf{q}^\times - \mathbf{q}^\times\hat{\varepsilon}\mathbf{q}^\times\mathbf{q}\hat{\varepsilon}\mathbf{q})$ is the reciprocal to the transposed matrix $\tilde{\hat{\varepsilon}}$, the sign $\circ$ designates the operation of the outer product of the vectors.

The integral propagation matrix $\hat{L}(d)$ for an ultrathin layer of thickness $d$ and $\hat{\chi}(z) = const$ can be easily calculated by the expansion of the matrix exponential up to the first order of $kd$:

$$\hat{L}(d) = \exp(ikd\hat{M}) = \begin{pmatrix} \hat{l}_1 & \hat{l}_2 \\ \hat{l}_3 & \hat{l}_4 \end{pmatrix} \approx 1 + ikd\hat{M} \approx$$

$$\approx \begin{pmatrix} 1 - ikd\chi_{yz}\cos\theta & 0 & ikd(1+\chi_{yy}) & -ikd\chi_{yx} \\ ikd\chi_{xz}\cos\theta & 1 & -ikd\chi_{xy} & ikd(\sin^2\theta + \chi_{xx}) \\ ikd\dfrac{\sin^2\theta + \chi_{zz}}{1+\chi_{zz}} & 0 & 1 - ikd\chi_{zy}\cos\theta & ikd\chi_{zx}\cos\theta \\ 0 & ikd & 0 & 1 \end{pmatrix}. \quad (26)$$

It is convenient to dismember the 4x4 matrix into 2x2 blocks $\hat{l}_j$, $j = 1,..4$.

The boundary task with the integral propagation matrix is presented by the system of the 4 equations (for the two two-dimensional vectors):

$$\begin{pmatrix} \mathbf{H}_t^T \\ \mathbf{q}\times\mathbf{E}^T \end{pmatrix} = \hat{L}(d)\begin{pmatrix} \mathbf{H}_t^0 + \mathbf{H}_t^R \\ \mathbf{q}\times\mathbf{E}^0 + \mathbf{q}\times\mathbf{E}^R \end{pmatrix} \quad (27)$$

We define the reflectivity amplitude $\hat{p}$ for the tangential components of the magnetic field of radiation by the relation:

$$\mathbf{H}_t^R = \hat{p}\,\mathbf{H}_t^0 \quad (28)$$

($\hat{p}$ is now a 2x2 matrix), and introduce the 2x2 matrices $\hat{\gamma}^{0,R,T}$, which supply the link between the tangential vectors $\mathbf{H}_t$ and $\mathbf{q}\times\mathbf{E}$ in the incident, reflected and transmitted waves (superscripts $0,R,T$ correspondingly)

$$\mathbf{q}\times\mathbf{E}^{0,R,T} = \hat{\gamma}^{0,R,T}\,\mathbf{H}_t^{0,R,T}. \quad (29)$$

Then the solution of (23) is presented by the following expression [51, 54]:

$$\hat{p} = [\hat{\gamma}^T(\hat{l}_1 + \hat{l}_2\hat{\gamma}^R) - (\hat{l}_3 + \hat{l}_4\hat{\gamma}^R)]^-[(\hat{l}_3 + \hat{l}_4\hat{\gamma}^0) - \hat{\gamma}^T(\hat{l}_1 + \hat{l}_2\hat{\gamma}^0)], \quad (30)$$

If a layer is in vacuum, the matrices $\hat{\gamma}^{0,R,T}$ has the very simple form:

$$\hat{\gamma}^{0,R,T} = \pm \begin{pmatrix} \sin\theta & 0 \\ 0 & \dfrac{1}{\sin\theta} \end{pmatrix}. \tag{31}$$

The calculation of the "numerator" in (26) in the first approximation gives

$$(\hat{l}_3 + \hat{l}_4 \hat{\gamma}^0) - \hat{\gamma}^T (\hat{l}_1 + \hat{l}_2 \hat{\gamma}^0) =$$

$$= \begin{pmatrix} ikd\dfrac{\sin^2\theta + \chi_{zz}}{1+\chi_{zz}} & 0 \\ 0 & ikd \end{pmatrix} + \begin{pmatrix} 1 - ikd\,\chi_{zy}\cos\theta & ikd\,\chi_{zx}\cos\theta \\ 0 & 1 \end{pmatrix} \begin{pmatrix} \sin\theta & 0 \\ 0 & \dfrac{1}{\sin\theta} \end{pmatrix} -$$

$$- \begin{pmatrix} \sin\theta & 0 \\ 0 & \dfrac{1}{\sin\theta} \end{pmatrix} [\begin{pmatrix} 1 - ikd\,\chi_{yz}\cos\theta & 0 \\ ikd\,\chi_{xz}\cos\theta & 1 \end{pmatrix} + \begin{pmatrix} ikd(1+\chi_{yy}) & -ikd\,\chi_{yx} \\ -ikd\,\chi_{xy} & ikd(\sin^2\theta + \chi_{xx}) \end{pmatrix} \begin{pmatrix} \sin\theta & 0 \\ 0 & \dfrac{1}{\sin\theta} \end{pmatrix}] =$$

$$= ikd \begin{pmatrix} \chi_{zz}\cos^2\theta - \chi_{yy}\sin^2\theta + (\chi_{yz} - \chi_{zy})\cos\theta\sin\theta & \dfrac{1}{\sin\theta}(\chi_{zx}\cos\theta + \chi_{yx}\sin\theta) \\ \dfrac{-1}{\sin\theta}(\chi_{xz}\cos\theta - \chi_{xy}\sin\theta) & -\chi_{xx}/\sin^2\theta \end{pmatrix}$$

$$\tag{32}$$

The calculation of the "denominator" in (26) in the first approximation gives

$$[\hat{\gamma}^T (\hat{l}_1 + \hat{l}_2 \hat{\gamma}^R) - (\hat{l}_3 + \hat{l}_4 \hat{\gamma}^R)] =$$

$$= \begin{pmatrix} 1 - ikd\,\chi_{zy}\cos\theta & ikd\,\chi_{zx}\cos\theta \\ 0 & 1 \end{pmatrix} \begin{pmatrix} \sin\theta & 0 \\ 0 & \dfrac{1}{\sin\theta} \end{pmatrix} - \begin{pmatrix} ikd\dfrac{\sin^2\theta + \chi_{zz}}{1+\chi_{zz}} & 0 \\ 0 & ikd \end{pmatrix} +$$

$$+ \begin{pmatrix} \sin\theta & 0 \\ 0 & \dfrac{1}{\sin\theta} \end{pmatrix} [\begin{pmatrix} 1 - ikd\,\chi_{yz}\cos\theta & 0 \\ ikd\,\chi_{xz}\cos\theta & 1 \end{pmatrix} - \begin{pmatrix} ikd(1+\chi_{yy}) & -ikd\,\chi_{yx} \\ -ikd\,\chi_{xy} & ikd(\sin^2\theta + \chi_{xx}) \end{pmatrix} \begin{pmatrix} \sin\theta & 0 \\ 0 & \dfrac{1}{\sin\theta} \end{pmatrix}] \cong$$

$$\cong \begin{pmatrix} 2\sin\theta & 0 \\ 0 & 2/\sin\theta \end{pmatrix}$$

$$\tag{33}$$

and

$$\begin{pmatrix} 2\sin\theta & 0 \\ 0 & 2/\sin\theta \end{pmatrix}^{-1} = \dfrac{1}{4}\begin{pmatrix} 2/\sin\theta & 0 \\ 0 & 2\sin\theta \end{pmatrix}. \tag{34}$$

Finally we get

$$\hat{p} = \frac{ikd}{2}\begin{pmatrix} 1/\sin\theta & 0 \\ 0 & \sin\theta \end{pmatrix} \times$$

$$\times \begin{pmatrix} \chi_{zz}\cos^2\theta - \chi_{yy}\sin^2\theta + (\chi_{yz} - \chi_{zy})\cos\theta\sin\theta & \frac{1}{\sin\theta}(\chi_{zx}\cos\theta + \chi_{yx}\sin\theta) \\ \frac{-1}{\sin\theta}(\chi_{xz}\cos\theta - \chi_{xy}\sin\theta) & -\chi_{xx}/\sin^2\theta \end{pmatrix} =$$

$$= \frac{ikd}{2\sin\theta}\begin{pmatrix} \chi_{zz}\cos^2\theta - \chi_{yy}\sin^2\theta + (\chi_{yz} - \chi_{zy})\cos\theta\sin\theta & \frac{1}{\sin\theta}(\chi_{zx}\cos\theta + \chi_{yx}\sin\theta) \\ -\sin\theta(\chi_{xz}\cos\theta - \chi_{xy}\sin\theta) & -\chi_{xx} \end{pmatrix}$$

(35)

In outer medium it is reasonable to have the reflectivity matrix $\hat{r}$ in $\sigma$- and $\pi$- polarization orts. If in the external medium $\chi_0 = 0$, the conversion from $\hat{p}$ to $\hat{r}$ is simple:

$$\hat{r} = \begin{pmatrix} r_{\sigma\sigma'} & r_{\sigma\pi'} \\ r_{\pi\sigma'} & r_{\pi\pi'} \end{pmatrix} = \begin{pmatrix} -p_{22} & p_{21}/\sin\theta \\ -p_{12}\sin\theta & p_{11} \end{pmatrix} \quad (36)$$

From (35), (36) we get the reflectivity matrix amplitude for a single ultrathin layer in the following form:

$$\hat{r} \cong \frac{ikd}{2\sin\theta}\hat{\chi}^\perp, \quad (37)$$

where we define the following transverse (to the propagation directions) susceptibility tensor $\hat{\chi}^\perp$ in $\sigma$- and $\pi$- polarization orts:

$$\hat{\chi}^\perp = \begin{pmatrix} \chi_{\sigma\to\sigma'} & \chi_{\pi\to\sigma'} \\ \chi_{\sigma\to\pi'} & \chi_{\pi\to\pi'} \end{pmatrix} =$$

$$= \begin{pmatrix} \chi_{xx} & \chi_{xy}\sin\theta - \chi_{xz}\cos\theta \\ -\chi_{yx}\sin\theta - \chi_{zx}\cos\theta & \chi_{zz}\cos^2\theta - \chi_{yy}\sin^2\theta + \cos\theta\sin\theta(\chi_{yz} - \chi_{zy}) \end{pmatrix}. \quad (38)$$

## IV. MAGNETIC CONTRIBUTION FROM A SINGLE LAYER

The magnetic contributions of the circular $\Delta\chi_m$ and linear $\Delta\chi_l$ dichroism to the susceptibility $\chi_0$ in the case of the dipole resonance transitions can be presented in the following form [49, 54]:

$$\hat{\chi} = \chi_0 + i\Delta\chi_m \mathbf{h}^\times + \Delta\chi_l \mathbf{h} \circ \mathbf{h}, \quad (39)$$

were $\mathbf{h}$ is the unit vector in the direction of magnetization, $\mathbf{h}^\times$ is the dual tensor:

$$\mathbf{h}^\times = \begin{pmatrix} 0 & -h_z & h_y \\ h_z & 0 & -h_x \\ -h_y & h_x & 0 \end{pmatrix}, \tag{40}$$

describing the operation of the vector product, the sign ∘ designates the operation of the outer product of the vectors, so $\mathbf{h} \circ \mathbf{h}$ is the tensor:

$$\mathbf{h} \circ \mathbf{h} = \begin{pmatrix} h_x h_x & h_x h_y & h_x h_z \\ h_y h_x & h_y h_y & h_y h_z \\ h_z h_x & h_z h_y & h_z h_z \end{pmatrix} \tag{41}$$

In general case the magnetization unit vector $\mathbf{h}$ has the following components (Fig. 2)

$$\mathbf{h} = (\sin\beta\cos\gamma, \sin\beta\sin\gamma, \cos\beta). \tag{42}$$

Typically $\Delta\chi_l$ is much smaller than $\Delta\chi_m$, so neglecting $\Delta\chi_l$ we have

$$\hat{\chi} = \chi_0 + \begin{pmatrix} 0 & -i\Delta\chi_m \cos\beta & i\Delta\chi_m \sin\beta\sin\gamma \\ i\Delta\chi_m \cos\beta & 0 & -i\Delta\chi_m \sin\beta\cos\gamma \\ -i\Delta\chi_m \sin\beta\sin\gamma & i\Delta\chi_m \sin\beta\cos\gamma & 0 \end{pmatrix} \tag{43}$$

Supposing that the magnetization in the ultrathin layer is arranged in the surface plane and along the grazing beam (L-MOKE geometry, $\beta=90°$ and $\gamma=90°$) we get:

$$\hat{\chi} = \chi_0 + \begin{pmatrix} 0 & 0 & i\Delta\chi_m \\ 0 & 0 & 0 \\ -i\Delta\chi_m & 0 & 0 \end{pmatrix} \tag{44}$$

and from (37), (38)

$$\hat{r} \cong \frac{ikd}{2\sin\theta} \begin{pmatrix} \chi_0 & -i\Delta\chi_m \cos\theta \\ i\Delta\chi_m \cos\theta & \chi_0 \cos 2\theta \end{pmatrix} \tag{45}$$

So for the σ-polarized incident wave, the amplitude of which we represent as $\begin{pmatrix} 1 \\ 0 \end{pmatrix}$, we have

$$r_{\sigma\to\sigma'} = \frac{ikd}{2\sin\theta}\chi_0 \quad \text{and} \quad r_{\sigma\to\pi'} = \frac{-kd}{2\sin\theta}\Delta\chi_m \cos\theta. \tag{46}$$

Taking use of (16), the reflectivity from such magnetic ultrathin layer placed under the reflecting substrate can be written:

$$\left|R^{tot}\right|^2 \approx \left|R^{substr}e^{iQ(H+d)} + (E_{tot}(H))^2 r_{\sigma\to\sigma'}\right|^2 + \left|(E_{tot}(H))^2 r_{\sigma\to\pi'}\right|^2 \quad (46)$$

where $E_{tot}(H) = 1 + R^{substr}e^{iQH}$ is the total radiation field amplitude at the position of the magnetic layer, $H$ is a distance of this layer from the substrate. The expression (40) contains two terms: the first one presents the reflectivity with the same polarization as the incident wave and the second one corresponds to the rotated $\pi$-polarization. It is clear from (46) that in the considered case this second part of the total reflectivity with the rotated $\pi$-polarization has the pure magnetic scattering origin:

$$\left|R_{\sigma\to\pi'}\right|^2 = \left|(E_{tot}(H))^2 \frac{kd\cos\theta}{2\sin\theta}\Delta\chi_m\right|^2. \quad (47)$$

In general case of the arbitrary magnetization direction the reflectivity with rotated polarization $\left|R_{\sigma\to\pi'}\right|^2$ takes the form:

$$\left|R_{\sigma\to\pi'}\right|^2 = \left|\frac{kd}{2\sin\theta}(E_{tot}(H))^2 \chi_{\sigma\to\pi'}\right|^2 =$$
$$= \left|\frac{kd}{2\sin\theta}(E_{tot}(H))^2 (-\chi_{yx}\sin\theta - \chi_{zx}\cos\theta)\right|^2 \quad (48)$$

Fig. 3 demonstrates the angular dependence of this dichroic component $\left|R_{\sigma\to\pi'}\right|^2$ for different distance $H$ of this thin layer from the substrate. The calculations are performed for the $L_2$ edge of gadolinium ($E_{ph}$=7930 eV). For the Gd layer of 1 Å thickness we put $\chi_0$=(-31.0+i*10.0)*$10^{-6}$ and $\Delta\chi_m$=(-0.1-i*0.23)*$10^{-6}$ (the data are taken form [67]), for the Si substrate $\chi_0$=(-15.6 +i 0.37) *$10^{-6}$ [68].

The obtained angular dependencies (Fig. 3) are similar to that ones presented in the paper [69] devoted to the standing waves influence on the fluorescent yield from heavy atoms incorporated into Langmuir layers. However, the contrast of oscillations in the calculated dependencies of the reflectivity with the rotated polarization is more pronounced due to the squared standing wave. It should have a consequence in the enhanced surface sensitivity. It is important that the curves in Fig. 3 calculated by the simple expression (47) and by the exact theory of the magnetic reflectivity with 4x4 propagation matrices [51, 54, 70] are identical.

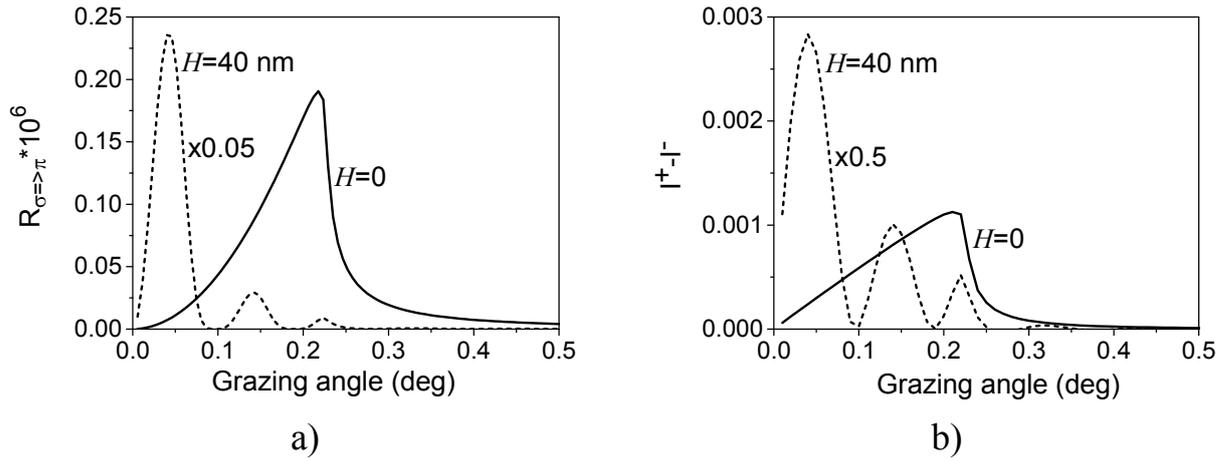

*Fig. 3. Angular dependences of the reflectivity with the rotated polarization (a) and the reflectivity difference $(I_+ - I_-)$ (b), calculated for the ultrathin magnetic layer, placed at different distance from Si substrate and magnetized along the beam. It is interesting to note, that in the considered case the peculiarities near the critical angle disappeared in the asymmetry angular dependence $(I_+ - I_-)/(I_+ + I_-)$ due to the strong variations of the denominator $(I_+ + I_-)$ (dot lines in (b)).*

It is reasonable to compare two possible kinds of measurements. The presented in Fig. 3a results need the linear polarization of the incident radiation and the polarization analysis of the reflected intensity. Commonly for magnetic investigations by the reflectivity method (XRMR) the circular polarization of the incident radiation is used and the asymmetry of the reflectivity curves over the signs of the used circular polarizations is analyzed. Supposing that the total reflectivity amplitude is a matrix in σ- and π- polarization orts

$$\hat{R}^{tot} = \begin{bmatrix} R_{\sigma\sigma} & R_{\sigma\pi} \\ R_{\pi\sigma} & R_{\pi\pi} \end{bmatrix}, \qquad (49)$$

the intensity of the reflected radiation for the right and left circular polarization $I_+$ and $I_-$ can be calculated according to this expressions:

$$I_+ = \frac{1}{2}(|R_{\sigma\sigma} + iR_{\sigma\pi}|^2 + |R_{\pi\sigma} + iR_{\pi\pi}|^2) =$$
$$= \frac{1}{2}\{|R_{\sigma\sigma}|^2 + |R_{\sigma\pi}|^2 + 2(\operatorname{Re} R_{\sigma\pi} \operatorname{Im} R_{\sigma\sigma} - \operatorname{Re} R_{\sigma\sigma} \operatorname{Im} R_{\sigma\pi}) + \qquad (50)$$
$$+ |R_{\pi\sigma}|^2 + |R_{\pi\pi}|^2 + 2(\operatorname{Re} R_{\pi\pi} \operatorname{Im} R_{\pi\sigma} - \operatorname{Re} R_{\pi\sigma} \operatorname{Im} R_{\pi\pi})\}$$

$$I_- = \frac{1}{2}(|R_{\sigma\sigma} - iR_{\sigma\pi}|^2 + |R_{\pi\sigma} - iR_{\pi\pi}|^2) =$$
$$= \frac{1}{2}\{|R_{\sigma\sigma}|^2 + |R_{\sigma\pi}|^2 - 2(\operatorname{Re} R_{\sigma\pi} \operatorname{Im} R_{\sigma\sigma} - \operatorname{Re} R_{\sigma\sigma} \operatorname{Im} R_{\sigma\pi}) + \tag{51}$$
$$+ |R_{\pi\sigma}|^2 + |R_{\pi\pi}|^2 - 2(\operatorname{Re} R_{\pi\pi} \operatorname{Im} R_{\pi\sigma} - \operatorname{Re} R_{\pi\sigma} \operatorname{Im} R_{\pi\pi})\}$$

and the difference in the reflectivity takes a form:

$$(I_+ - I_-) = 2(\operatorname{Re} R_{\pi\pi} \operatorname{Im} R_{\pi\sigma} - \operatorname{Re} R_{\sigma\sigma} \operatorname{Im} R_{\sigma\pi} +$$
$$+ \operatorname{Im} R_{\sigma\sigma} \operatorname{Re} R_{\sigma\pi} - \operatorname{Im} R_{\pi\pi} \operatorname{Re} R_{\pi\sigma}) \tag{52}$$

From (51) it follows that the reflectivity amplitudes with the rotated polarization, which have purely magnetic scattering origin and typically very small, are enhanced in this XRMR method by the much higher $R_{\sigma\sigma}$ and $R_{\pi\pi}$ reflectivity amplitudes of scattering without polarization change. This circumstance makes the measurements of the magnetic scattering easier, but on the other hand it essentially complicates the data treatment directed at the extraction of the magnetic scattering information. The expression (52) disproves the assertion of the authors in [71, 72] about a pure magnetic origin of the measured asymmetry ratio. Note that the selection of the purely magnetic scattering part had been done in [73] by a complicated combination of different kinds of measurements (L-MOKE, T-MOKE and Faraday rotation) on a ferromagnetic Fe/C multilayer at the Fe-2p absorption edge.

It is clear that using the selection of the reflectivity with the rotated $\pi$-polarization by a polarization analyzer directly gives the purely magnetic scattering part separately from the dominant $R_{\sigma\sigma}$ reflectivity.

## IV. MAGNETIC REFLECTIVITY FROM THE WHOLE MAGNETIC STRUCTURE

If magnetic scattering is small enough and does not influence on the total radiation field inside the whole sample, we can suppose that the magnetic scattering from different layers are independent from each other. In this case we can summarize these magnetic scattering amplitudes with proper phases like it is done in the kinematical theory of reflectivity. The analogous procedure has been used in [38]. However, contrary to [38] we will take into account the influence of the variations of the radiation field $E(\theta,z)$ at different depths $z$. So, for the calculation of the reflectivity with the rotated polarization we take use of (48) and suggest the following expression:

$$I_{\sigma \to \pi'}(\theta) = \left| \frac{\pi}{\lambda \sin\theta} \int \chi^{\sigma \to \pi'}(z) E_\sigma^{\,2}(\theta, z)\, dz \right|^2, \qquad (53)$$

where $\chi^{\sigma \to \pi'}$ is the magnetic off-diagonal elements of the transverse susceptibility tensor $\hat{\chi}^{\perp}$ (38).

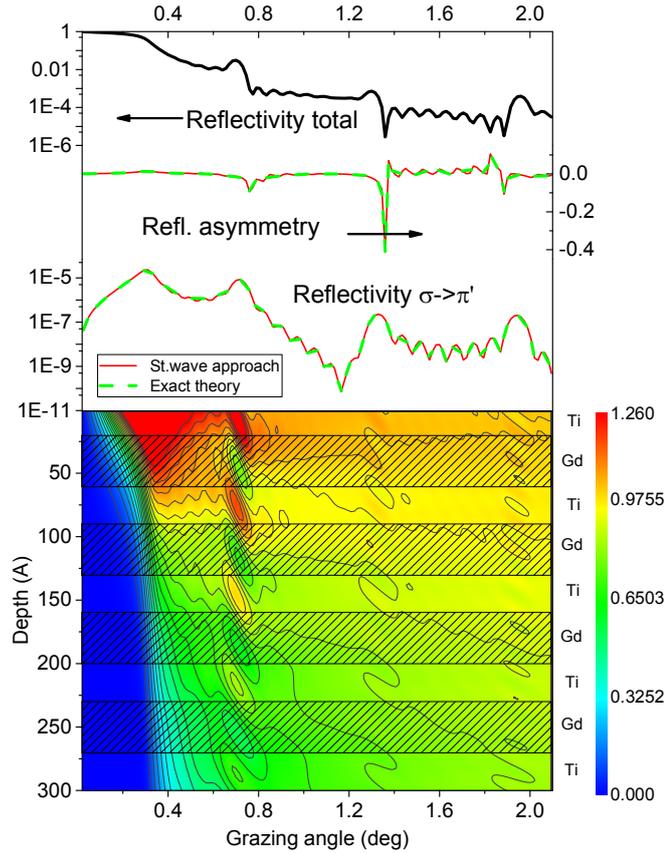

*Fig. 4. Squared standing waves $\left|E_\sigma^{\,2}(\theta,z)\right|^2$ inside [Ti(3 nm)/Gd(4 nm)]$_8$ multilayer (bottom part, colored on-line), the rotated reflectivity and asymmetry ratio, calculated by the exact theory and by (53), (54) (middle part of the picture) and the total reflectivity (top graph). The magnetic contributions to the reflectivity originates only from Gd layers (hatched), magnetization in which is supposed ferromagnetically ordered along the beam (L-MOKE geometry). The calculations for $E_{ph}=7930$ eV and with the same parameters of Gd susceptibility as in Fig. 3.*

The calculation results presented in Fig. 4 demonstrate that such improved kinematical approximation (53) gives the angular curves of the reflectivity with the rotated polarization perfectly reproducing the exact calculations [70] for all angles of incidence including the region of the total external reflection.

The calculated total field depth-distribution (more precisely squared standing waves) drown in Fig. 4 is independent on the type of the magnetization ordering in multilayer and it can be used as the basis for the calculations of the rotated reflectivity and asymmetry ratio by (53), (52), (54) for ferromagnetic, antiferromagnetic and spiral interlayer coupling between Gd layers. The calculation results of these cases are shown in Fig. 5.

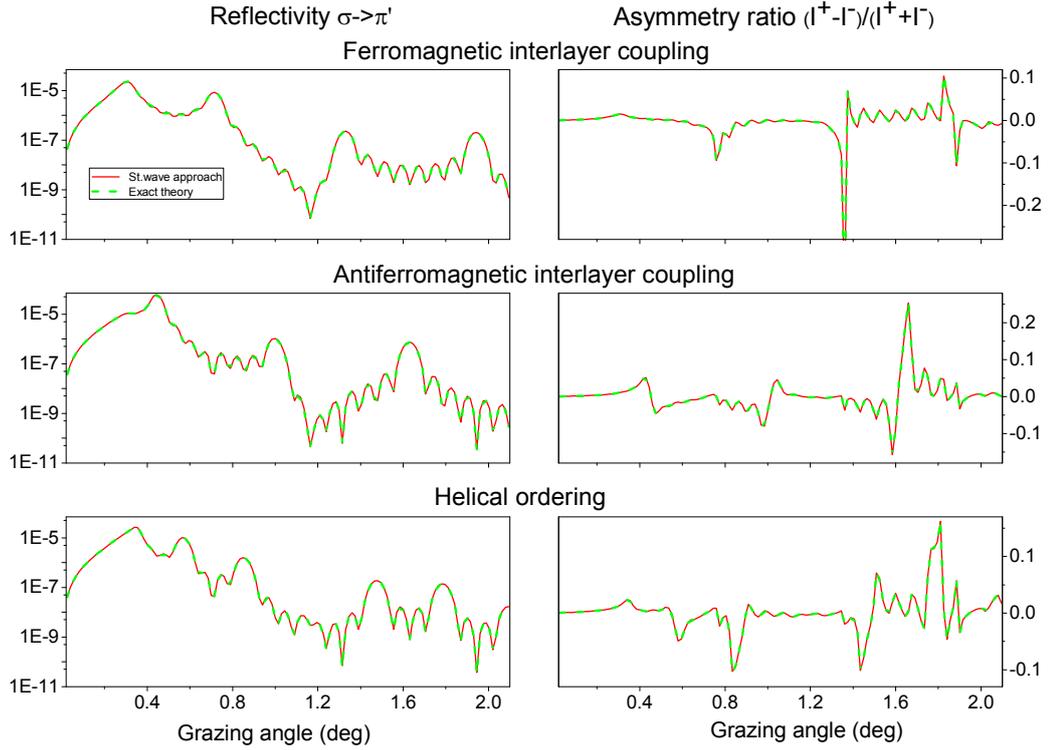

*Fig. 5. Angular dependencies of the rotated reflectivity (left part) and asymmetry ratio (right part), calculated by the exact theory and by (53), (54) for different cases of the magnetic ordering between Gd layers: for ferromagnetic interlayer coupling [Gd(4 nm)/Ti(3 nm)]$_8$, for antiferromagnetic interlayer coupling [Gd↑(4 nm)/Ti(3 nm)/Gd↓(4 nm)/Ti(3 nm)]$_4$ and for helical ordering with magnetic period 28 nm.*

For the asymmetry ratio calculations by the expression (52) the off-diagonal elements of the amplitude reflectivity matrix can be obtained similarly to (53) (but without squared module):

$$R_{\sigma\pi(\pi\sigma)}(\theta) = \frac{i\pi}{\lambda \sin\theta} \int \chi^{\sigma\pi(\pi\sigma)}(z) E_{\sigma(\pi)}^2(\theta,z)\, dz \ . \qquad (54)$$

Note that in most cases $R_{\pi\sigma} = R_{\sigma\pi}$ and $E_\sigma(\theta,z) = E_\pi(\theta,z)$. The diagonal elements of the amplitude reflectivity matrix $R_{\sigma\sigma}$ and $R_{\pi\pi}$ in (52) can be calculated ignoring the magnetic contribution to the susceptibilities of the layers by the simple e.g. Parratt algorithm. Figs. 4,5 show that such way of the asymmetry ratio calculation gives as well the results identical to the exact calculations [70].

Figs. 4,5 show that both the angular curves for the asymmetry ratio and for the reflectivity with rotated polarization characterize the peculiarities of the magnetic ordering by specific maxima at Bragg angles or the magnetic satellites, but the shapes of the two dependences are rather different. Therefore, it can be supposed that the additional to the asymmetry ratio measurements of the rotated polarization can give an additional information that is important for the investigations of the complicated magnetic structures.

The significance of the expression (53) is stipulated by two aspects. Firstly, it allows us to calculate the magnetic reflectivity much faster without 4x4 propagation matrices, because the most complicated part of the standing wave calculations can be done by the scalar theory of reflectivity (by Parratt algorithm). It essentially speeds-up the model calculations and fit procedure. Secondly, the interpretation of the reflectivity with the rotated polarization based on the squared standing waves explains the exclusive depth selectivity of the measurements using the polarization analysis.

### V. EXPERIMENTAL TEST

The most important evidence of the standing wave influence on the magnetic reflectivity with rotated polarization is presented by the peak near the critical angle of the total reflection. The first experimental test of this peculiarity of the angular curve of reflectivity with the rotated polarization has been done for the Ti(10 nm)/Gd$_{0.23}$Co$_{0.77}$(250 nm)/Ti(10 nm) film at the L$_2$ edge of Gd [74]. The sample has the compensation temperature of T$_{comp}\approx$433 K for Co and Gd subsystem magnetizations [75], so at room temperature the Gd atoms should possess the magnetic moments. The measurements were performed at the Kurchatov Center for Synchrotron Radiation and Nanotechnology (KCSRN). The sample was placed on the piece of the permanent magnet in order to magnetize it along the beam. The analysis of the polarization state of the reflected radiation was performed by 90º-reflection from Si crystal ((422) reflection with 2θ$_B$ =89.682º for λ=0.156 nm) placed before the detector. The peak for the reflectivity with rotated polarization at the critical angle was observed (Fig. 6) only for the resonant photon energy $E_{ph}$=7930 eV. So we can attribute its origin to the magnetic scattering on the Gd atoms. The very small value of the "dichroic" effect was explained not only by the small concentration of Gd atoms in the film, but as well by the rather thick Ti top layer, preventing the penetration of the incident radiation to the $Gd_{0.23}Co_{0.77}$ layer. So the standing wave created in the angular region in vicinity of the critical angle could excite only the Gd atoms in the Ti/$Gd_{0.23}Co_{0.77}$ interface.

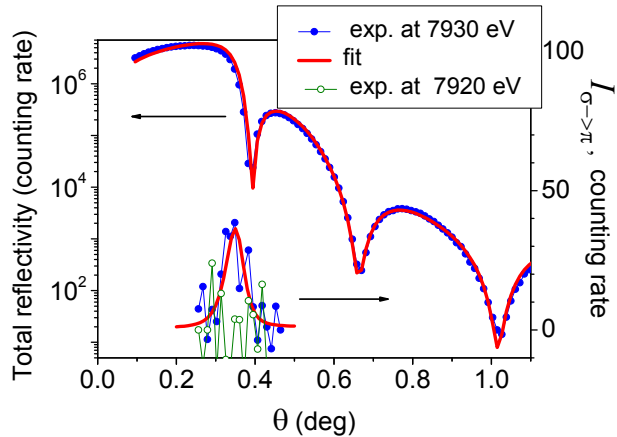

*Fig. 6. Experimental total reflectivity (the left log scale) and reflectivity with the rotated polarization (the right normal scale) from Ti(10 nm)/$Gd_{0.23}Co_{0.77}$(250 nm)/Ti(10 nm) multilayer for $E_{ph}$=7930 eV (filled symbols, blue on-line) and $E_{ph}$=7920 эB (open symbols, green on=line). (For details see [71]).*

Mössbauer scattering on the $^{57}$Fe containing samples is also characterized by the dichroic effect for some hyperfine components of the spectrum. In order to get the angular dependencies of the nuclear resonant reflectivity (NRR) the integral over reflectivity resonant spectrum is measured as the function of the grazing incidence angle of the beam. The measurements of the NRR curve has been done at ID18, ESRF [76], using the Synchrotron Mössbauer Source (SMS [77, 78]) for the sample [$^{57}$Fe(0.8 nm)/Cr(1.05 nm)]$_{30}$ at 4 K and external field of 5 T [79] (in order to align the magnetizations in 57Fe layers ferromagnetically).

The radiation from SMS is purely $\pi$-polarized, and the rotated polarization should have the $\sigma$-polarization state. The polarization analysis of the reflected $\sigma$-polarized radiation was done by the Si channel-cut crystal (two (840) reflection with $2\theta_B$ =90.2$^o$ for $\lambda$=0.086 nm). Unfortunately, the sample surface was not good enough and the reflected beam has a rather broad angular distribution (~200"), so the integration over angular scans with the Si channel-cut crystal at each angle of incidence was needed.

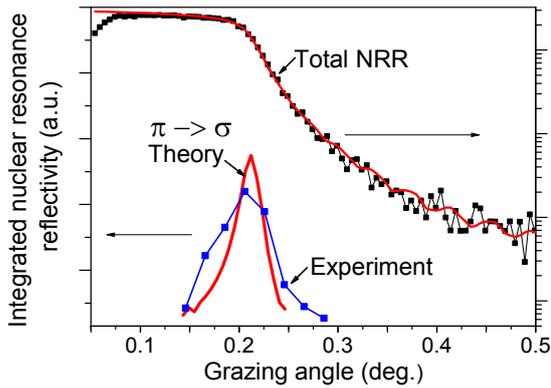

FIG. 7. *NRR angular curves measured without polarization selection (logarithmic right scale) and with the selection of the σ-polarized reflectivity (normal left scale).*

The result is shown in Fig. 7. Again we observe the peak at the critical angle, confirming the standing wave influence on the weak magnetic scattering with rotated polarization.

## IV. SUMMARY

We have deduced the formula (53), (54) describing the small magnetic contribution to the x-ray reflectivity in the kinematical approximation which is valid at all angles including the total reflection region. The approach takes into account the typically small value of the x-ray magnetic scattering amplitude and based on the exact calculations of the radiation field amplitude inside the reflecting multilayer. We found that the squared standing waves (forth power of the radiation field amplitude) determine the magnetic scattering at each depth. From the described formalism the appearance of the peak near the critical angle of the total reflection should appear for the selected part of the reflectivity (with rotated polarization). The first experimental test confirms this prediction. The model calculations show the full agreement of the results obtained by the described approach with the exact calculations.

There are several points stipulating the significance of the developed approach. Firstly, it can essentially speed-up the calculations of the x-ray resonant magnetic reflectivity. Secondly, the connection of the rotated polarization part in the reflectivity with the standing waves inside the medium reveals the enhanced depth selectivity of the polarization analysis in the resonant reflectivity measurements. And finally, as it was pointed in [45, 46], for the subpicosecond time domain measurements, developing nowadays on the platform of the x-ray free electron laser facilities, the polarization effects in the reflectivity (e.g. Kerr rotation) is becoming increasingly important.

## ACKNOWLEDGMENTS

This work was supported by the Russian Foundation of the Basic Research through the Grants No. 15-02-1502-a and 16-02-00887-a.

___________________________________________________________________

1. Ch. Brouder, J. Phys.: Condens. Matter. **2**, 701-738 (1990).
2. G. Schütz, V. Knülle, and H. Ebert, In: "Resonant anomalous x-ray scattering", Ed.by G. Materlik, C.J. Sparks and K. Fischer, Elsevier Science, 1994.
3. J. Stöhr, Journal of Electron Spectroscopy and Related Phenomena **75**, 253-272 (1995).
4. C.R. Natoli, Ch. Brouder, Ph. Sainctavit, J. Goulon, Ch. Goulon-Ginet, and A. Rogalev, Eur. Phys. J. **B** 4, 1-11 (1998).
5. J. Goulon, A. Rogalev, F. Wilhelm, C. Goulon-Ginet, P. Carra, I. Marri, Ch. Brouder, Journal of Experimental and Theoretical Physics **97**, 402-431 (2003).
6. G. van der Laan, Journal of Physics: Conference Series **430**, 012127 (2013).
7. R. Sessoli, M.-E. Boulon, A. Caneschi1, M. Mannini1, L. Poggini1, F. Wilhelm, and A. Rogalev, Nature physics, Nature Physics **11**, 69–74 (2014).
8. D.E. Moncton, D. Gibbs, J. Bohr, Nuclear Instruments and Methods in Physics Research A **246**, 839-844 (1986).
9. M. Blume, D. Gibbs, Phys. Rev. B **37**, 1779-1789 (1988).
10. D**.** Gibbs, D. R. Harshman, E. D. Isaacs, D. B. McWhan, D. Mills, C. Vettier, Phys. Rev. Lett. **61**, 1241–1244 (1988).
11. J. Bohr, D. Gibbs, J.D. Axe, D.E. Moncton, K. L. D'Amico, C. F. Majkrzak, J. Kwo. M. Hong, C.L. Chien, J. Jensen, Physica B **159,** 93-105, (1989).
12. D. B. McWhan, C. Vettier, E. D. Isaacs, G. E. Ice, D. P. Siddons, J. B. Hastings, C. Peters. O. Vogt, Phys. Rev. B **42**, 6007–6017 (1990).
13. J. Bohr, Journal of Magnetism and Magnetic Materials **83**, 530-534 (1990).
14. T. Bruckel, M. Lippert, T. Köhler, J. R. Schneider, W. Prandl, V. Rilling, and M. Sschilling, Acta Cryst. A **52**, 427-437 (1996).

15. S. Langridge and G. H. Lander, N. Bernhoeft, A. Stunault, C. Vettier, G. Grübel, C. Sutter, F. de Bergevin, W. J. Nuttall, W. G. Stirling, K. Mattenberger, and O. Vogt, Phys. Rev. B **55**, 6392–6398 (1997).
16. V. Fernandez and C. Vettier, F. de Bergevin, C. Giles, W. Neubeck, Phys. Rev. B **57,** 7870 (1998).
17. W. Neubeck, C. Vettier, V. Fernandez, F. De Bergevin, and C. Giles, J. of Applied Physics **85**, 4847-4849 (1999).
18. D. P. Siddons, M. Hart, Y. Amemiya, J. B. Hastings, Phys. Rev. Lett. **64**, 1967-1970 (1990).
19. S P Collins, J. Phys.: Condens. Matter. **11**, 1159–1175 (1999).
20. H.-Ch. Mertins, F. Schäfers, A. Gaupp, X. Le Cann, W. Gudat, Phys. Rev. B **61**, R874-877 (2000).
21. J. B. Kortright, Sang-Koog Kim, Phys. Rev. B **62**, 12 216 (2000).
22. H.-Ch. Mertins, P. M. Oppeneer, J. Kuneš, A. Gaupp, D. Abramsohn, F. Schäfers, Phys. Rev. Lett. **87**, 047401-1-4 (2001).
23. R. Caciuffo, L. Paolasini, A. Sollier, P. Ghigna, E. Pavarini, J. van den Brink*,* and M. Altarelli, Phys. Rev. B **65**, 174425 (2002).
24. Manabu Takahashi, Manabu Usuda, and Jun-ichi Igarashi, Phys. Rev. B **67**, 064425 (2003).
25. G. Beutier, S. P. Collins, O. V. Dimitrova, V. E. Dmitrienko, M. I. Katsnelson, Y. O. Kvashnin, A. I. Lichtenstein, V. V. Mazurenko, A. G. A. Nisbet, E. N. Ovchinnikova, and D. Pincini, Phys. Rev. Lett. **119**, 167201 (2017).
26. C. C. Kao, C. T. Chen, E. D. Johnson, E. D. Hastings, H. J. Lin, G. H. Ho, G. Meigs, J. M. Brot, S. L. Hulbert, Y. U. Idzerda, and C. Vettier, Phys. Rev. B **50**, 9599 (1994).
27. J. M. Tonnerre, L. Sève, D. Raoux, G. Soullié, B. Rodmacq, and P. Wolfers, Phys. Rev. Lett. **75**, 740 (1995).
28. M. Gibert, M. Viret1, P. Zubko, N. Jaouen, J.-M. Tonnerre, A. Torres-Pardo, S. Catalano, A. Gloter, O. Sterphan, and J.-M. Triscone, Nature communications 7, 11227 (2016).
29. S. Brück, G. Schütz, E. Goering, X. Ji, and K. M. Krishnan, Phys. Rev. Lett. **101**, 126402 (2008).
30. J. Geissler, E. Goering, M. Justen, F. Weigand, and G. Schűtz, J. Langer, D. Schmitz, H. Maletta, and R. Mattheis, Phys. Rev. B **65**, 020405 R (2001).
31. N. Ishimatsu, H. Hashizume, S. Hamada, N. Hosoito, C. S. Nelson, C. T. Venkataraman, G. Srajer, and J. C. Lang, Phys. Rev. B **60**, 9596 (1999).

32. L. Sève, N. Jaouen, J. M. Tonnerre, D. Raoux, F. Bartolomé, M. Arend, W. Felsch, A. Rogalev, J. Goulon, C. Gautier, and J. F. Bérar, Phys. Rev. B **60**, 9662 (1999).
33. N. Jaouen, G. van der Laan, T. K. Johal, F. Wilhelm, A. Rogalev, S. Mylonas, and L. Ortega, Phys. Rev. B **70**, 094417 (2004).
34. N. Hosoito, T. Ohkochi, K. Kodama, and R. Yamagishi, Journal of the Physical Society of Japan **78**, 094716 (2009).
35. S.-K. Kim, and J. B. Kortright, Phys. Rev. Lett. **86**, 1347 (2001).
36. S. M. Valvidares, C. Quirós, A. Mirone, J.-M. Tonnerre, S. Stanescu, P. Bencok, Y. Souche, L. Zárate, J. I. Martín, M. Vélez, N. B. Brookes, and J. M. Alameda, Phy. Rev. B **78**, 064406 (2008).
37. M. Elzo, R. Moubah, C. Blouzon, M. Sacchi, S. Grenier, R. Belkhou, S. Dhesi, D. Colson, F. Torres, M. Kiwi, M. Viret, and N. Jaouen, Phys. Rev. B **91**, 014402 (2015).
38. H. Ott, C. Schubetaler-Langeheine, E. Schierle, G. Kaindl, E. Weschke, Appl. Phys. Lett. **88**, 212507 (2006).
39. A Bergmann, J Grabis, A Nefedov, K Westerholt and H Zabel, J. Phys. D: Appl. Phys. **39**, 842–850 (2006).
40. J. W. Freeland, J. Chakhalian, A. V. Boris, J.-M. Tonnerre, J. J. Kavich, P. Yordanov, S. Grenier, P. Zschack, E. Karapetrova, P. Popovich, H. N. Lee, and B. Keimer, Phys. Rev. B **81**, 094414 (2010).
41. J. Kortright, and M. Rice, Rev. Sci. Instrum. **67**, 3353 (1996).
42. P. Oppeneer, Magneto-optical Kerr spectra. Handb. Magn. Mater. **13**, 229–422 (2001).
43. H. C. Mertins, S. Valencia, D. Abramsohn, A. Gaupp, W. Gudat, and P. M. Oppeneer, Phys. Rev. B **69**, 064407 (2004).
44. R. Azzam and N. Bashara, Ellipsometry and Polarized Light, North-Holland P.C., 1977.
45. Sh. Yamamoto, M. Taguchi, T. Someya, Y. Kubota, S. Ito, H. Wadati, M. Fujisawa, F. Capotondi, E. Pedersoli, M. Manfredda, L. Raimondi, M. Kiskinova, J. Fujii, P. Moras, T. Tsuyama, T. Nakamura, T. Kato, T. Higashide, S. Iwata, S. Yamamoto, S. Shin, and I. Matsuda, Review of Scientific Instruments **86**, 083901 (2015).
46. Sh. Yamamoto, and I. Matsuda, Appl. Sci. **7**, 662 (2017).
47. J. Zak, E. R. Moog, C. Liu, and S. D. Bader, Phys. Rev. B **43**, 6423 (1991).
48. A. Bourzami, O. Lenoble, Ch. Fe'ry, J. F. Bobo, and M. Piecuch, Phys. Rev. **59**, 11489 (1999).

49. S. A. Stepanov and S. A. Sinha, Phys. Rev. B **61**, 15302 (2000).
50. M. Elzo, E. Jal, O. Bunau, S. Grenier, Y. Joly, A. Y. Ramos, H. C. N. Tolentino, J. M. Tonnerre, and N. Jaouen, J. Magn. Magn. Mater. **324**, 105 (2012).
51. L. M. Barkovskii, G. N. Borzdov, and V. Lavrukovich, Zh. Prikladnoi Spektroskopii **25**, 526 (1976).
52. 21. L. M. Barkovskii, G. N. Borzdov, and F. I. Fedorov, Wave Operators in Optics, Preprint of Inst. of Physics, Belarusian Acad. Sci., Minsk, 1983, No. 304.
53. M. A. Andreeva and C. Rosete, Vestnik Moskovskogo Universiteta, Fizika **41** (3), 57-62 (1986).
54. M. Andreeva, and A. Smekhova, Applied Surface Science **252**, 5619 (2006).
55. M. Andreeva, A. Smekhova, B. Lindgren, M. Bjorck, and G. Anderson, Journal of Magnetism and Magnetic Materials **300**, e371 (2006).
56. N. K. Pleshanov, Z. Phys. B: Condens. Matter **94**, 233 (1994).
57. A. Ruhm, B. P. Toperverg, and H. Dosch, S Phys. Rev. B **60**, 16073 (1999).
58. E. Kravtsov, D. Haskel, S. G. E. te Velthuis, J. S. Jiang, and B. J. Kirby, Phys. Rev. B **79**, 1334438 (2009).
59. E. E. Odintsova, and M. A. Andreeva, Journal of Surface Investigation. X_ray, Synchrotron and Neutron Techniques **4**, No. 6, 913–922 (2010).
60. M.A. Andreeva, Yu. L. Repchenko, Crystallography Reports **58**(7), 1037 (2013).
61. M. Born, and E.Wolf, Principles of optics, Pergamon Press, 1968.
62. L. G. Parratt, Phys. Rev. **95**, 359 (1954).
63. Toellner, T. L., Sturhahn, W., Röhlsberger, R., Alp, E. E., Sowers, E. E. & Fullerton, E. E. (1995), Phys. Rev. Lett. **74**, 3475-3478.
64. A. Q. R. Baron, J. Arthur, S. L. Ruby, A. I. Chumakov, G. V. Smirnov, and G. S. Brown, Phys. Rev. B **50**, 10354-10357 (1994).
65. M. A. Andreeva, and B. Lindgren, JETP Letters **76**(12), 704-706 (2002).
66. M.A. Andreeva, and B. Lindgren, Phys. Rev. B **72**, 125422-1-22 (2005).
67. C. Sorg, A. Scherz, K. Baberschke, H. Wende, F. Wilhelm and A. Rogalev, S. Chadov, J. Minár, and H. Ebert, Phys. Rev. B **75**, 064428-1-5 (2007).
68. http://henke.lbl.gov/optical_constants/getdb2.html
69. M. J. Bedzyk, G. M. Bommarito, J. S. Schildkraut, Phys. Rev. Lett. **62**, 1376 (1989).
70. M. A. Andreeva, Yu. L. Repchenko, http://kftt.phys.msu.ru/index.php?id=47 (X_ray_magn_Refl).

71. H. Höchst, D. Zhao, and D. L. Huber, Surf. Sci. **352–354**, 998-1002 (1996).
72. H. Höchst, D. Rioux, D. Zhao, and D. L. Huber, J. Appl. Phys. **81**, 7584–7588 (1997).
73. H.-Ch. Mertins, D. Abramsohn, A. Gaupp, F. Schäfers, W. Gudat, O. Zaharko and H. Grimmer,P. M. Oppeneer, Phys. Rev. B **66**, 184404 (2002).
74. M. A. Andreeva, R. A. Baulin , M. M. Borisov, E. A. Gan'shina, G. V. Kurlyandskaya, E. Kh. Mukhamedzhanov, Yu. L. Repchenko, A. V. Svalov, Journal of Experimental and Theoretical Physics, 126(5), in press (2018).
75. A. V. Svalov, G. V. Kurlyandskaya, K. G. Balymov, V. O. Vas'kovskii, Spin valves based on amorphous ferrimagnetic Gd–Co films, The Physics of Metals and Metallography **117** (9), 876-882 (2016).
76. R. Rüffer, R. and A. I. Chumakov, Hyperfine Interactions **97/98**, 589-604 (1996).
77. V. Potapkin, A. I. Chumakov, G. V. Smirnov, R. Rüffer, C. McCammon, and L. Dubrovinsky, Phys. Rev. A **86**, 053808 (2012).
78. V. Potapkin, A. I. Chumakov, G. V. Smirnov, J.-P. Celse, R. Rüffer, C. McCammon, and L. Dubrovinsky, J. Synchrotron Rad. **19**, 559-569 (2012).
79. M. A. Andreeva, R. A. Baulin, A. I. Chumakov, and R. Rüffer, Journal of Applied Physics, submitted; arXiv 1803.10117.